\documentclass[notitlepage,superscriptaddress,nofootinbib,tightenlines,preprintnumbers,twocolumn,floatfix]{revtex4-2}
\usepackage{amsmath,verbatim,latexsym,amssymb,indentfirst,mathrsfs,mathtools,amsthm,bbm,bm, url,cancel,subcaption,enumitem}
\usepackage{natbib}

\usepackage[font=small,labelfont=bf,format=plain,justification=raggedright,singlelinecheck=false]{caption}
\usepackage{verbatim,indentfirst}
\usepackage[title,titletoc]{appendix}

\usepackage{lineno} % For numbering the lines
\usepackage[colorlinks]{hyperref}
\hypersetup{%
	plainpages=true,
	breaklinks=true,% not default in dvips mode, so we must specify
	hypertexnames=false,%not ideal, but needed when pagenums duplicate (`i' vs. `1')
	pageanchor=true,
	colorlinks=true,
	linkcolor={blue},
	citecolor={red},
	urlcolor={blue},
	anchorcolor={black}
}

\usepackage{graphicx}
\usepackage{epstopdf}
\renewcommand{\thesection}{\Roman{section}}
\renewcommand{\thesubsection}{\Roman{section} \Alph{subsection}}
\renewcommand{\thesubsubsection}{\Roman{section} \Alph{subsection} \arabic{subsubsection}}
\makeatletter
\def\p@subsection{}
\makeatother
\makeatletter
\def\p@subsubsection{}
\makeatother

\makeatletter
\newcommand\footnoteref[1]{\protected@xdef\@thefnmark{\ref{#1}}\@footnotemark}
\makeatother

\usepackage{soul}

\usepackage{color}
\usepackage[dvipsnames]{xcolor}
\usepackage[normalem]{ulem}

\newcommand{\1}{{\rm H}}

\def\id{\mathbbm{1}}   % Identity
\newcommand*{\bra}[1]{\langle #1\rvert}
\newcommand*{\ket}[1]{\lvert #1 \rangle}

\begin{document}
 
\title{Thermally driven quantum refrigerator autonomously resets superconducting qubit}

%\simone{Autonomous, thermally-driven reset of a superconducting qubit based on a quantum absorption refrigerator}

%
\author{Mohammed Ali Aamir}
\email{aamir.ali@chalmers.se}
\affiliation{%
Department of Microtechnology and Nanoscience, Chalmers University of Technology, 412 96 Gothenburg, Sweden
}
\author{Paul Jamet Suria}
\affiliation{%
Department of Microtechnology and Nanoscience, Chalmers University of Technology, 412 96 Gothenburg, Sweden
}
\author{Jos\'e Antonio Mar\'in Guzm\'an}
\affiliation{Joint Center for Quantum Information and Computer Science, NIST and University of Maryland, College Park, MD 20742, USA}
\author{Claudia Castillo-Moreno}
\affiliation{%
Department of Microtechnology and Nanoscience, Chalmers University of Technology, 412 96 Gothenburg, Sweden
}
\author{Jeffrey M. Epstein}
\author{Nicole~Yunger~Halpern}
\email{nicoleyh@umd.edu}
\affiliation{Joint Center for Quantum Information and Computer Science, NIST and University of Maryland, College Park, MD 20742, USA}
\affiliation{Institute for Physical Science and Technology, University of Maryland, College Park, MD 20742, USA}
\author{Simone Gasparinetti}
\email{simoneg@chalmers.se}
\affiliation{%
Department of Microtechnology and Nanoscience, Chalmers University of Technology, 412 96 Gothenburg, Sweden
}
\date{\today}

%
% Abstract
%
\begin{abstract}

Although classical thermal machines power industries and modern living, quantum thermal engines have yet to prove their utility. Here, we demonstrate a useful quantum absorption refrigerator formed from superconducting circuits. We use it to cool a transmon qubit to a temperature lower than that achievable with any one available bath, thereby resetting the qubit to an initial state suitable for quantum computing. The process is driven by a thermal gradient and is autonomous, requiring no external feedback. The refrigerator exploits an engineered three-body interaction between the target qubit and two auxiliary qudits. Each auxiliary qudit is coupled to a physical heat bath, realized with a microwave waveguide populated with synthesized quasithermal radiation. If the target qubit is initially fully excited, its effective temperature reaches a steady-state level of approximately 22~mK, lower than what can be achieved by existing state-of-the-art reset protocols. Our results demonstrate that superconducting circuits with propagating thermal fields can be used to experimentally explore quantum thermodynamics and apply it to quantum information-processing tasks.

\end{abstract}

\maketitle
    
%\linenumbers

\begin{figure*}
\includegraphics[width=1\linewidth]{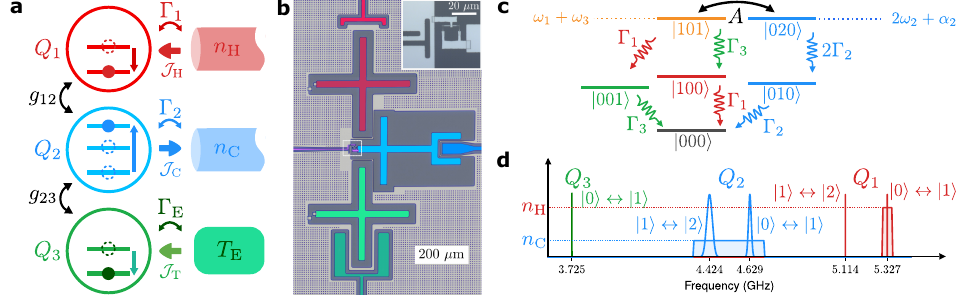}
\caption{Quantum-absorption-refrigerator scheme and level diagram. \textbf{a,}~Conceptual scheme with three qudits. Qubit $Q_1$ couples directly to a waveguide at a rate $\Gamma_1$; and qudit $Q_2$, to another waveguide at a rate $\Gamma_2$. Qubit $Q_3$ couples undesirably to an uncontrolled bath in its environment. This bath keeps $Q_3$ at an effective temperature $T_{\rm E}$. The coupling rate $\Gamma_{\rm relax}$ determines $Q_3$'s natural energy-relaxation time. The waveguides can operate as heat baths containing photons of average numbers $n_{\rm H}$ and $n_{\rm C}$. The interqudit couplings ($g_{12}$, $g_{23}$) engender a process in which one excitation in $Q_1$ and one excitation in $Q_3$ are simultaneously exchanged with a double excitation in $Q_2$. This exchange helps reset $Q_3$. When heat baths drive this process, the system operates as an autonomous quantum refrigerator. The average heat currents are depicted by wide arrows from the hot ($\mathcal{J}_\mathrm{H}$), cold ($\mathcal{J}_\mathrm{C}$), and target ($\mathcal{J}_\mathrm{T}$) systems' baths. By energy conservation, $\mathcal{J}_\mathrm{H} + \mathcal{J}_\mathrm{T} = \mathcal{J}_\mathrm{C}$. \textbf{b,}~False-color micrograph of the device implemented with superconducting circuits. $Q_2$ is frequency-tunable, due to a flux current line and two parallel Josephson junctions, magnified in the inset. \textbf{c,}~Level diagram showing tensor products $\ket{q_1 q_2 q_3}$ of the qudits' energy eigenstates. $\ket{101}$ and $\ket{020}$ are resonant if 
$\omega_1 + \omega_3 = 2\omega_2 + \alpha_2$. 
At resonance, a three-body interaction couples the states at a rate $A$. 
\textbf{d,}~Distributions over the qudits' experimentally observed transition frequencies. The Lorentzian distributions' widths represent spectral widths. 
The red shaded box depicts the spectral density $n_{\rm H}$ of photons injected into the waveguide coupled to $Q_1$. This synthesized noise realizes the refrigerator's hot thermal bath. Analogous statements concern the blue box, $n_{\rm C}$, $Q_2$, and the cold bath.
}\label{Fig1}
\end{figure*}

Quantum thermodynamics should be more useful. The field has yielded fundamental insights, such as extensions of the second law of thermodynamics to small, coherent, and far-from-equilibrium systems~\cite{lieb1999, janzing2000, egloff2015, horodecki2013, brandao2015, yungerhalpern2016, halpern2018, lostaglio2017, guryanova2016, yungerhalpern2016a, sparaciari2017, gour2018a, khanian2023}. Additionally, quantum phenomena have been shown to enhance engines~\cite{kim2011c, gelbwaser-klimovsky2018, myers2020, lostaglio2020, kalaee2021, hammam2022}, batteries~\cite{binder2015}, and refrigeration~\cite{jennings2010, levy2020}. These results are progressing gradually from theory to proof-of-principle experiments. However, quantum thermal technologies remain experimental curiosities, not practical everyday tools. Key challenges include control~\cite{woods2023} and cooling quantum thermal machines to temperatures that support quantum phenomena. 
Both challenges require substantial energy and effort but yield small returns. For example, one would expect a single-atom engine to perform only about an electronvolt of work~\cite{rossnagel2016}.

Autonomous quantum machines offer hope. First, they operate without external control. Second, they run on heat drawn from thermal baths, which are naturally abundant~\cite{mitchison2015}. A quantum thermal machine would be useful in a context that met three criteria: (i) The machine fulfills a need. (ii) The machine can access real-world different-temperature baths. (iii) No or few extra resources are spent on maintaining whatever coherence is necessary for the machine's operation.

We identify such a context: qubit reset. Consider a superconducting quantum computer starting a calculation. The computer requires qubits initialized to their ground states~\cite{divincenzo2000a}. If left to thermalize with its environment as thoroughly as possible, though, a qubit could achieve only an excited-state population of $\approx$0.01 to 0.03, or an effective temperature of 45~mK to 70~mK~\cite{geerlings2013, jin2015a, magnard2018, zhou2021}. Furthermore, such passive thermalization takes a few multiples of the qubit's energy-relaxation time---hundreds of microseconds in state-of-the-art setups---delaying the next computation. A quantum machine cooling the qubits to their ground (minimal-entropy) states fulfills criterion (i). Moreover, superconducting qubits inhabit a dilution refrigerator formed from nested plates, whose temperatures decrease from the outermost plate to the innermost. These temperature plates can serve as heat baths, meeting criterion (ii). Finally, the machine can retain its quantum nature if mounted on the coldest plate, next to the quantum processing unit, satisfying criterion (iii). Such an autonomous machine would be a \emph{quantum absorption refrigerator}.

Quantum absorption refrigerators have been widely studied theoretically~\cite{palao2001, linden2010, levy2012,  chen2012g, venturelli2013, correa2014, silva2015, hofer2016a, mu2017, nimmrichter2017, du2018, holubec2019, mitchison2018, mitchison2019, naseem2020, manikandan2020, arrangoiz-arriola2018, bhandari2021, kloc2021, almasri2022, okane2022}. Reference~\cite{maslennikov2019} reported a landmark proof-of-principle experiment performed with trapped ions. However, the heat baths were emulated with electric fields and lasers, rather than realized with physical heat reservoirs. Other quantum refrigerators, motivated by possible applications, have been proposed~\cite{manikandan2023, karmakar2022a} and tested~\cite{baugh2005, solfanelli2022, buffoni2023} but are not autonomous. 

We report on a quantum absorption refrigerator realized with superconducting circuits. Our quantum refrigerator cools---and therefore resets---a target superconducting qubit \emph{autonomously}. The target qubit’s energy-relaxation time is fully determined by the temperature of a hot bath we can vary. Using this control, we can vary the energy-relaxation time by a factor of $>70$. The reset's fidelity is competitive: The target's excited-state population reaches below $3 \times 10^{-4} \pm 2 \times 10^{-4}$ [effective temperatures as low as $22~(+2,-3)$~mK]. 
In comparison, state-of-the-art reset protocols achieve populations of $8\times 10^{-4}$ to $2 \times 10^{-3}$ (effective temperatures of 40~mK to 49~mK)~\cite{magnard2018,zhou2021}. Our experiment demonstrates that quantum thermal machines not only can be useful, but also can be integrated with quantum information-processing units. Furthermore, such a practical autonomous quantum machine costs less control and thermodynamic work than its nonautonomous counterparts~\cite{ronzani2018, klatzow2019, tan2017b}. 

Our absorption refrigerator consists of three qudits ($d$-level quantum systems), as depicted in Fig.~\ref{Fig1}a. The auxiliary qudits $Q_1$ and $Q_2$ correspond to $d=2,3$, respectively. Each of them couples directly to a waveguide that supports a continuum of electromagnetic modes. The waveguide can serve as a heat bath formed from photons of an arbitrary spectral profile. $n_{\rm H}$ and $n_{\rm C}$ denote the average numbers of photons in the waveguides. The target of the refrigerator's cooling is qubit $Q_3$, which is undesirably coupled to an uncontrolled bath in its environment. This bath excites the target to an effective temperature $T_{\rm E}$. Nearest-neighbor qudits couple together with strengths $g_{12}$ and $g_{23}$. These couplings result in an effective three-body interaction~\cite{ren2020}, a crucial ingredient in a quantum absorption refrigerator~\cite{linden2010, levy2012, mitchison2015, hofer2016a}. We engineer the three-body interaction such that one excitation in $Q_1$ and one excitation in $Q_3$ are simultaneously, coherently exchanged with a double excitation in $Q_2$. Losing its excitation, $Q_3$ is reset.

As the heat baths drive the resetting, the system operates autonomously as a quantum absorption refrigerator~\cite{mitchison2019}. A generic thermodynamic model describes such a refrigerator as follows. Heat flows from a hot bath (coupled to $Q_1$) into an intermediate-temperature bath (coupled to $Q_2$). A heat current $\mathcal{J}_\mathrm{H}$ ($\mathcal{J}_\mathrm{T}$) flows out of $Q_1$'s ($Q_3$'s) bath. A net heat current $\mathcal{J}_\mathrm{C} = \mathcal{J}_\mathrm{H} + \mathcal{J}_\mathrm{T}$ enters $Q_2$'s bath (Fig.~\ref{Fig1}a). That is, a temperature gradient, rather than work, coaxes heat out of the target qubit.

The qudits are Al-based superconducting transmons that have Al/AlO$_\mathrm{x}$/Al Josephson junctions~\cite{koch2007a}. We arrange the qudits spatially in a linear configuration  (Fig.~\ref{Fig1}b). The capacitances between the transmons couple the qudits mutually. Qudit $Q_1$ has a transition frequency $\omega_1/(2\pi) = 5.327$~GHz; and qudit $Q_2$, a variable frequency $\omega_2/(2\pi)$. $Q_1$ couples capacitively to a microwave waveguide directly, at a dissipation rate $\Gamma_1/(2\pi) = 70$~kHz; and $Q_2$ couples to another waveguide at $\Gamma_2/(2\pi) = 7.2$~MHz. The third qubit, $Q_3$, has a transition frequency $\omega_3/(2\pi) = 3.725$~GHz. $Q_3$ couples dispersively to a coplanar waveguide resonator. Via the resonator, we read out $Q_3$'s state and drive $Q_3$ coherently. In addition, $Q_3$ couples to the uncontrolled bath in its environment at a rate $\Gamma_{\rm E}$. In our proof-of-concept demonstration, $Q_3$ stands in for a computational qubit that is being reset and that may participate in a larger processing unit. In the present design, $Q_3$ has a natural energy-relaxation time $T_{\rm relax} = 1/\Gamma_{\rm E} = 16.8~\mu$s, limited largely by Purcell decay into the nearest waveguide, and a residual excited-state population $P_{\rm res}=0.028$. In future realizations, one can increase $T_{\rm relax}$ using Purcell filters~\cite{reed2010}.

The interqudit couplings hybridize the qudit modes. The hybridization, together with the Josephson junctions' nonlinearity, results in a three-body interaction (Supplementary Information). For this interaction to be resonant, the qudit frequencies must meet the condition $\omega_1 + \omega_3 = 2\omega_2^{{\rm res}} + \alpha_2$. Here, $\omega_2^{{\rm res}}$ denotes the $Q_2$ frequency that satisfies the equality, and $\alpha_2$ denotes $Q_2$'s anharmonicity. The interaction arises from a four-wave mixing process: One excitation in $Q_1$ and one excitation in $Q_3$ are simultaneously exchanged with a double excitation in $Q_2$ (Fig.~\ref{Fig1}a)~\cite{ren2020}. To satisfy the resonance condition \emph{in situ}, we make $Q_2$ frequency-tunable~\cite{koch2007a}. We control the frequency with a magnetic flux induced by a nearby current line. The device is mounted in a dilution refrigerator that reaches 10~mK.

To describe the resonance condition, we introduce further notation. Let $\ket{0}$ and $\ket{1}$ denote the ground and first-excited states of any qudit. Let $\ket{2}$ denote the second-excited state of  $Q_2$. We represent a three-qudit state by $\ket{q_1 q_2 q_3} := \ket{q_1}_1 \otimes \ket{q_2}_2 \otimes \ket{q_3}_3$. The resonance condition leads to coherence between the states $|101\rangle$ and $|020\rangle$. This coherence is a key quantum feature of our refrigerator. Two processes, operating in conjunction, reset $Q_3$: (i) Levels $\ket{101}$ and $\ket{020}$ coherently couple with an effective strength $A$ (Fig.~\ref{Fig1}c). 
(ii) $Q_2$ dissipates into its waveguide at a rate $\Gamma_2$. The combined action of (i) and (ii) brings $\ket{101}$ rapidly to $\ket{010}$ (and then to $\ket{000}$), thereby resetting $Q_3$.

\begin{figure}[h]
\includegraphics[width=0.99\linewidth]{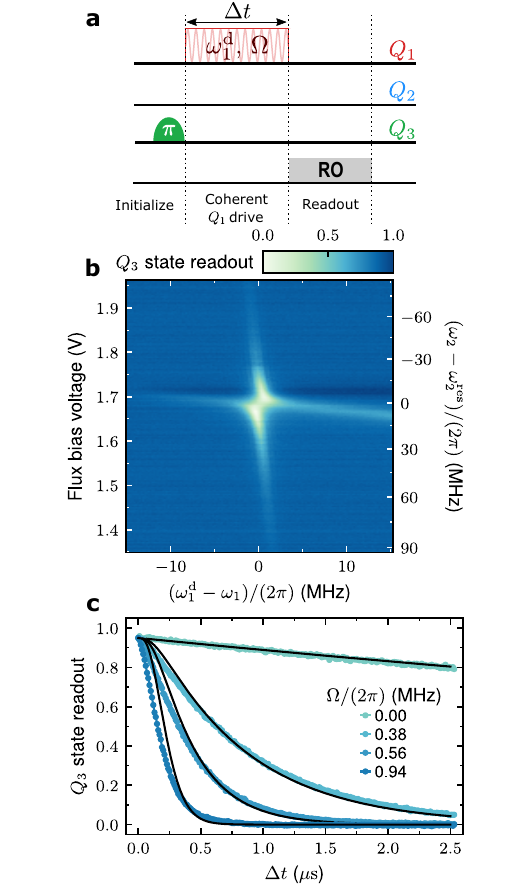}
\caption{Three-body interaction: \textbf{a,}~Pulse scheme (see the main text for description). 
\textbf{b,}~2D plot of $Q_3$'s excited-state population ($[\langle\sigma_z\rangle + \id] /2$), as a function of (i) the flux current (left axis) modulating $Q_2$'s frequency and (ii) the detuning between the $Q_1$-drive frequency $\omega_1^{\rm d}/(2\pi)$ and $\omega_1/(2\pi)$ (bottom axis). $Q_1$ is driven for $\Delta t = 2~\mu$s during the pulse scheme, after which we read out (RO) $Q_3$'s state via the resonator. The left axis translates directly into the right axis---the detuning of the $Q_2$ frequency, $\omega_2/(2\pi)$, from the resonant value, $\omega_2^{{\rm res}}/(2\pi)$. 
The white patch evidences an avoided crossing, where $\ket{101}$ and $\ket{020}$ become resonant (Fig.~\ref{Fig1}c). \textbf{c,}~Excited-state readout of $Q_3$ as a function of the duration $\Delta t$ of the $Q_1$ drive, at select drive rates $\Omega/(2\pi)$. The solid black lines are fits based on the model shown in Supplementary Section~II.}\label{Fig2}
\end{figure}

We engineer the heat baths of $Q_1$ and $Q_2$ as follows. First, we synthesize radiation using room-temperature electronics (Supplementary Figure~S1). This radiation has a white-noise spectral profile over a selected frequency range. The radiation is injected into microwave coaxial cables, which are interlinked by dissipative microwave attenuators thermalized at different temperatures of the cryostat. The attenuators reduce the incoming radiation's power, while simultaneously introducing quantum thermal noise~\cite{caves1982, fink2010, scigliuzzo2020}. The last attenuator, at 10~mK, contributes noise that is predominantly quantum vacuum noise. This resulting radiation finally reaches $Q_1$'s and $Q_2$'s waveguides. Quantum noise is generally characterized by a non-symmetrical spectral density resulting in different emission and absorption rates of a qudit~\cite{clerk2010a}. Here, the predominance of spontaneous emission of $Q_2$ into the cold bath (as opposed to absorption) is critical to the refrigerator's operation. 

\begin{figure}[h]
\includegraphics[width=1\linewidth]{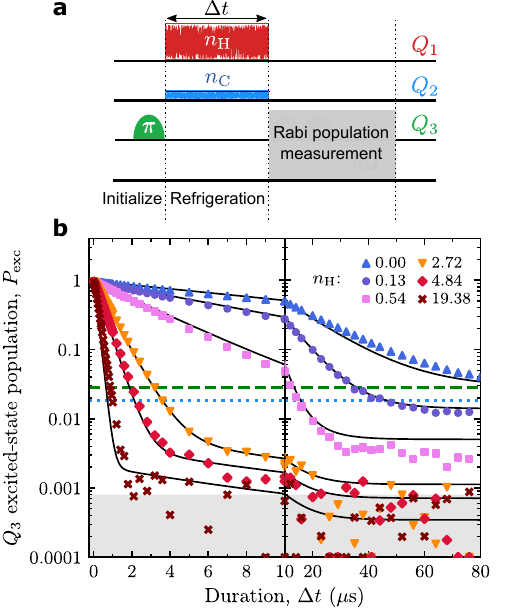}
\caption{Autonomous refrigeration enabled by a hot bath. \textbf{a,}~Three-step pulse scheme: Initialization brings $Q_3$'s state close to $\ket{1}$. During refrigeration, $Q_1$ and $Q_2$ interact with synthesized quasithermal fields for a duration $\Delta t$. Finally, $Q_3$'s excited-state population, $P_{\rm exc}$, is measured via a Rabi population-measurement scheme. $P_{\rm exc}$ represents combined populations of the first and second excited states; the latter is calculated based on a fitted theoretical model and is negligible except at intermediate values of $\Delta t$ (Supplementary Section~III). \textbf{b,}~$P_{\rm exc}$ as a function of $\Delta t$, at select values of $n_{\rm H}$, the average number of photons in the hot bath. The $x$-axis is split into two \emph{different} regimes of $\Delta t$ values: a low regime $\Delta t \in [0, 10]$ $\mu$s and a high regime $\Delta t \in [10, 80]$ $\mu$s. %The $x$-axis's scale differs between the two regimes. 
$Q_2$ experiences no synthesized quasithermal field. We estimate that $n_{\rm C} \approx 0.007$ due to the residual thermal field. The dashed green line shows $Q_3$'s residual excited-state population (defined in the main text), $P_{\rm res} = 0.028$. The dotted blue line shows the excited-state population that $Q_3$ would have at the cold bath's temperature (45~mK), $P_{\rm C} = 0.020$. The gray area represents our estimate of the noise floor (Supplementary Section~IV). Near the noise floor, some measurements yield small negative values, represented by data points at the bottom axis. Solid lines represent global fits to the experimental curves. The fits are calculated from the model shown in Supplementary Section~II. $n_{\rm H}$ is the sole free-fitting parameter.
}\label{Fig3}
\end{figure} 

The bandwidth of the synthesized radiation is selected to include the frequencies of $Q_1$'s transitions ($|0\rangle \leftrightarrow |1\rangle$) and $Q_2$'s transitions ($|0\rangle \leftrightarrow |1\rangle$ and $|1\rangle \leftrightarrow |2\rangle$) (Fig.~\ref{Fig1}d). Within this bandwidth, the radiation can be approximated as a thermal field. Outside the bandwidth, however, the radiation deviates from thermality. Therefore, we designate this field as \emph{quasithermal}. Its effective temperature, $T_{\rm H,C}$, depends on the average number of photons at the qudit $Q_{1,2}$ transition frequency: $n_\mathrm{H, C} = 1/[\exp(\hbar \omega_{1, 2}/k_{\rm B} T_{{\rm H,C}}) - 1]$. $k_{\rm B}$ is Boltzmann's constant. We can vary $n_\mathrm{\rm H, C}$ by regulating the synthesized noise's power. This setup enables the whole system to function as a quantum thermal machine. The quasithermal baths induce transitions in $Q_1$ and $Q_2$, autonomously driving the reset via the three-body interaction.

Having specified the setup, we demonstrate the three-body interaction: We verify that $Q_3$ can be reset via resonant driving of $Q_1$ if and only if $Q_2$ meets the resonance condition. The qudits begin in $\ket{000}$, whereupon we issue two microwave drive pulses (Fig.~\ref{Fig2}a). The first is a Gaussian $\pi$-pulse that excites $Q_3$ to state $|1\rangle$: $\ket{000} \rightarrow \ket{001}$.

The second pulse is flat and coherently drives $Q_1$ (effecting $\ket{001} \leftrightarrow \ket{101}$) at a frequency $\omega_1^{\rm d}$, with a rate $\Omega$, for a duration $\Delta t$. Subsequently, we perform qubit-state readout on $Q_3$ (we measure $[\langle \sigma_z \rangle + \id] / 2$) via $Q_3$'s resonator. 

We investigate the readout's dependences on $\omega_1^{\rm d}$ and on the flux bias voltage (proportional to flux current) that modulates the tunable $Q_2$ frequency. We have fixed $\Delta t = 2~\mu$s and $\Omega/(2\pi) = 200$~kHz. The microwave drives, we observe, deplete $Q_3$'s excited-state population (Fig.~\ref{Fig2}a). The depletion is the greatest when $\omega_2 = \omega_2^{\mathrm{res}}$ and the drive is resonant ($\omega_1^{\rm d} = \omega_1$)---when the resonant coupling $A$ between $|101\rangle$ and $|020\rangle$ is the strongest. 

The excited state of $Q_3$ is depleted by the cascaded processes $\ket{001} \leftrightarrow \ket{101} \leftrightarrow \ket{020} \rightarrow \ket{010}$. The combined effect of these processes resembles optical pumping---used to achieve population inversion in atomic physics---enabling qubit reset. Away from the resonance condition, the resonant coupling $A$ decreases. Consequently, the excited-state population of $Q_3$ drops less as the $\ket{101}$--$\ket{020}$ detuning grows. 

Furthermore, we study the effect of increasing the driving rate $\Omega$ (Fig.~\ref{Fig2}c). When $\Omega = 0$~MHz, $Q_3$ decays to its ground state (resets) at its natural energy-relaxation time ($16.8~\mu$s). As $\Omega$ increases, the reset happens increasingly quickly. By fitting a model based on a Lindblad master equation (Supplementary Section~II), we determine that the three-body interaction has a strength of $A/(2\pi) = 3.2$~MHz. 

Having demonstrated the three-body interaction, we operate the three-qubit system as a quantum thermal machine. To measure the system's performance, we implement a three-step pulse sequence (Fig.~\ref{Fig3}a): (1) Excite $Q_3$ to near $|1\rangle$ (to an excited-state population of 0.95). (2) Fill the waveguides with quasithermal photons, as described above, for a variable time interval $\Delta t$. (3) Measure $Q_3$'s excited-state population (the first two excited states combined), using a Rabi population-measurement scheme~\cite{geerlings2013, jin2015a}. This scheme allows for a more accurate population measurement than standard qubit-state readout. This scheme functions optimally when $Q_3$'s second-excited-state population is negligible compared to the first-excited-state population. However, this condition may not always be met when the latter is extremely small ($\lesssim 0.004$). Nonetheless, we account for $Q_3$'s second-excited-state population, determined theoretically from a comprehensively fitted model, in all population measurements, which are recalibrated accordingly (Supplementary Section~III). We assume that the second-excited-state population is exponentially suppressed, arising from the same uncontrolled bath causing the residual first-excited-state population. This recalibration is insignificant except in some narrow subsets of our experimental data, which lie outside the regime in which we evaluate our refrigerator's performance (see below).

We raise the effective temperature of the hot bath and investigate how $P_{\rm exc}$ responds. To do so, we elevate the average number $n_{\rm H}$ of quasithermal photons in the hot bath by increasing the spectral power of the synthesized noise in $Q_1$'s waveguide. We perform this study in the absence of synthesized noise in the cold bath (coupled to $Q_2$), which contains the minimal average number $n_{\rm C}$ of photons. We infer the minimal $n_{\rm C}$ from an independent measurement, using $Q_2$ as a thermometer~\cite{scigliuzzo2020}: $n_{\rm C} = 0.007$, associated with a temperature $T_{\rm C} = 45$~mK. The greater the $n_{\rm H}$ value, the more quickly $P_{\rm exc}$ decays as we increase $\Delta t$ (Fig.~\ref{Fig3}b). At the low value $n_{\rm H} = 0.16$, $P_{\rm exc}$ drops below the residual excited-state population $P_{\rm res} = 0.028$ (green dashed line in Fig.~\ref{Fig3}b), which $Q_3$ would achieve if left alone for a long time. We infer from this value that the effective temperature of $Q_3$'s environment bath $T_{\rm E} = 50$~mK (see Fig.~1a). If thermalized at the cold bath's temperature (45~mK), $Q_3$ would have an excited-state population $P_\mathrm{C} = 0.020$ (blue dotted line). If the hot bath is excited, $P_{\rm exc}$ reaches at least an order of magnitude lower than $P_{\rm res}$ and $P_{\rm C}$. Our refrigeration scheme clearly outperforms passive thermalization with either the intrinsic bath of $Q_3$ or the coldest bath available. At $n_{\rm H} = 19.38$ ($T_{\rm H} = 5.1$~K), refrigeration reduces the effective energy-relaxation time of $Q_3$, $T_{\rm relax}$, from $16.8~\mu$s to $230$~ns. This reduction is by a factor of $> 70$. $Q_3$'s population declines below $2\times 10^{-3}$ over $1.8~\mu$s, before approaching a steady-state value below $0.0008$, our measurement protocol's noise floor (Supplementary Section~IV).

\begin{figure}
\includegraphics[width=1\linewidth]{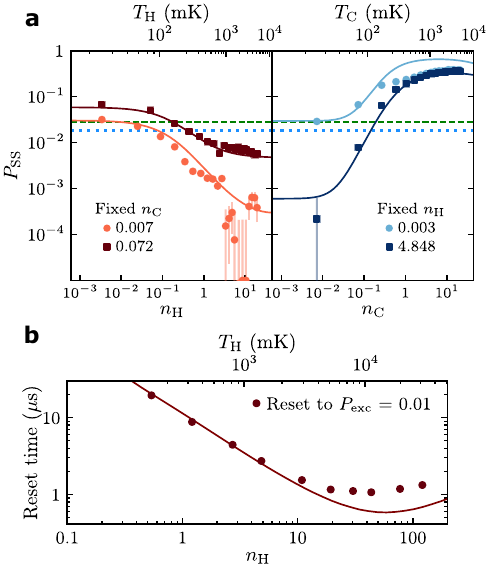}
\caption{Performance metrics of the quantum absorption refrigerator. \textbf{a,}~After a 105~$\mu$s reset protocol, $Q_3$'s excited state reaches a steady-state excited-state population $P_{\rm SS}$.~(Left) $P_{\rm SS}$ as a function of the hot bath’s average photon number, $n_{\rm H}$. The corresponding temperature $T_{\rm H}$ is translated along the top axis. Two experimental curves are at two values of the cold bath’s average photon number, $n_{\rm C}$. (Right) $P_{\rm SS}$ as a function of $n_{\rm C}$ (translated into a temperature $T_{\rm C}$ on the top axis), at two $n_{\rm H}$ values. Some measurements yield small negative values, represented by data points at the bottom axis. The dashed green and dotted blue lines are the same as in Fig.~3b. \textbf{b,}~Reset time (time required for $Q_3$’s $P_{\rm exc}$ to reach 0.01), as a function of $n_{\rm H}$. All solid lines are theoretical predictions calculated from the model shown in Supplementary Section~II.
}\label{Fig4}
\end{figure}

In an independent measurement, we study the steady-state population $P_{\rm SS}$ as a function of $n_{\rm H}$ or $n_{\rm C}$, keeping the other quantity fixed (Fig.~\ref{Fig4}a). We define $P_{\rm SS}$ as the $P_{\rm exc}$ achieved after the duration $\Delta t = 105~\mu\mathrm{s}$. This definition stems from the observation that, when the refrigerator is inactive ($n_\mathrm{H} = 0.003$), $Q_3$ naturally relaxes to its steady-state residual population, $P_{\rm res}$, by $\Delta t = 105~\mu$s. $P_{\rm SS}$ decreases rapidly as $n_{\rm H}$ increases. Furthermore, $P_{\rm SS}$ reaches its lowest values when $n_{\rm C}$ minimizes at $0.007$, such that $Q_2$ is not excited. We overestimate the lowest reached $P_{\rm SS}$ and its error margin by computing the mean and standard deviation of all the measured $P_{\rm SS}$ values that lie below $0.0008$, the noise floor (Supplementary Section~IV shows the methodology). $P_{\rm SS}$ reaches a minimum $< 3 \times 10^{-4} \pm 2 \times 10^{-4}$, equivalent to a temperature $T_{\rm SS} = 22~(+2,-3)$~mK. This result is remarkably close to the prediction from a general theory of a quantum absorption refrigerator~\cite{mitchison2019}: $T_{\rm SS} = \frac{2\omega_2+\alpha_2 - \omega_1}{(2\omega_2 +\alpha_2)/T_{\rm C} - \omega_1/T_{\rm H}} = 18.6$~mK, equivalent to $P_{\rm SS} = 6.7\times 10^{-5}$. In the limit as $n_{\rm H} \rightarrow \infty$, $T_{\rm SS}$ decreases marginally to $18.5$~mK. $T_\mathrm{SS}$ does not depend on the temperature $T_{\rm E}$ of the target's effective bath, if $\Gamma_{\rm E}$ is very small ($ \ll 1/T_{\rm relax}$), as in our system during refrigeration.

Also, raising the cold bath's temperature impedes the reset. Increasing $n_{\rm C}$ to 0.07---exciting $Q_2$ more---leads $P_{\rm SS}$ (as a function of $n_{\rm H}$) to saturate at a higher value. Finally, consider fixing $n_{\rm H}$ and increasing $n_{\rm C}$. $P_{\rm SS}$ increases rapidly, then saturates near 0.36. This saturation occurs largely independently of $n_{\rm H}$. The greater the $n_{\rm H}$, though, the greater the initial (low-$n_{\rm C}$) $P_{\rm SS}$. %As $n_{\rm C}$ decreases, $T_{\rm SS}$ decreases further.

A standard figure of merit in the thermodynamic analysis of refrigerators is the coefficient of performance (COP)~\cite{hofer2016a}. The COP is to refrigerators as efficiency is to heat engines. The steady-state COP is defined as $\mathcal{J}_\mathrm{T}/\mathcal{J}_\mathrm{H}$ (see Fig.~1a), that we numerically calculate from the theoretical model shown in Supplementary Section~II. The steady-state COP is 0.7 when $T_{\rm H} = 5.1$~K and $T_{\rm SS} = 22$~mK. In terms of COP, our quantum refrigerator performs comparably to a macroscopic absorption refrigerator---namely, a common air conditioner (COP $\approx 0.7$~\cite{CIBSE}). In the quasistatic limit (as $\mathcal{J}_\mathrm{T} \rightarrow 0$), the COP can reach its theoretical upper bound, the Carnot limit: $\frac{T_{\rm E}(T_{\rm H} - T_{\rm C}) }{ T_{\rm H}(T_{\rm C} - T_{\rm E})}$. Our quantum refrigerator has a Carnot bound of $0.95 > 0.7$, satisfying the second law of thermodynamics.

Another important performance metric is the time required to reset $Q_3$. We define the reset time as the time required for $P_{\rm exc}$ to reach $0.01$ (corresponding to $38.5$~mK). The reset time reaches as low as 970~ns before rising slowly with $n_{\rm H}$ (Fig.~\ref{Fig4}b). We attribute the observed upturn to excessive dephasing of the coherent process $|101\rangle \leftrightarrow |020\rangle$, which is critical for refrigeration.

%\section{Conclusion}

In summary, we have demonstrated the first quantum thermal machine being deployed to accomplish a useful task. The task---reset of a superconducting qubit---is crucial to quantum information processing. The machine---a quantum absorption refrigerator formed from superconducting circuits---cools and resets the target qubit to an excited-state population lower than that achieved with state-of-the-art active reset protocols, without requiring external control. Nevertheless, the refrigeration can be turned off when the target qubit serves in a computation: one can either change the hot bath's temperature or detune a qudit out of resonance, using an on-chip magnetic flux.

Our refrigerator has two main quantum features---discrete energy levels and a coherent exchange coupling between states $|101\rangle$ and $|020\rangle$. Another salient feature of our quantum thermal machine is its use of waveguides as physical heat baths. In contrast, other experiments have emulated heat baths~\cite{klatzow2019, maslennikov2019}. Our heat baths consist of quasithermal fields---syntheses of quantum thermal fields and finite-bandwidth artificial microwave noise. Our approach allows control over the baths' temperatures, the ability to tailor spectral properties of the heat baths, and the selection of the level transitions to be heated. Thus, this method can facilitate a rigorous study of quantum thermal machines. Our experimental setup can be modified to exploit real-world thermal baths, such as different-temperature plates of a dilution refrigerator. We have already demonstrated that our quantum refrigerator can reset a qubit effectively if it has access to a hot bath at a temperature of a few kelvin, without the need for tuning. Superconducting coaxial cables, together with infrared-blocking filters \cite{rehammar2023}, can expose the qudits to thermal radiation emitted by hot resistors anchored to the suitable dilution refrigerator's plate \cite{goetz2017, wang2021}. The modification adds no significant heat load to the base-temperature plate; nor does it compromise the performance of the quantum information-processing unit. One can activate the thermal reset on demand in two different ways: (1) by using a microwave switch~\cite{pechal2016} to toggle $Q_1$'s bath between hot and cold or (2) by dynamically detuning $Q_2$ in and out of the resonance condition that enables the reset process.

Our quantum refrigerator initiates a path toward experimental studies of quantum thermodynamics with superconducting circuits coupled to propagating thermal microwave fields. Superconducting circuits may also offer an avenue toward scaling quantum thermal machines similarly to quantum-information processors. Our experiment may inspire the further development of useful, real-world applications of quantum thermodynamics~\cite{guzman2023} to quantum information processing~\cite{fellous-asiani2021, auffeves2022, aifer2022}, thermometry~\cite{mehboudi2019, scigliuzzo2020, wang2021}, algorithmic cooling~\cite{baugh2005, alhambra2019}, timekeeping~\cite{erker2017}, and entanglement generation~\cite{brask2015}. This work marks a significant step in quantum thermodynamics toward practical applications.

\begin{appendices}

% To switch from two-column to one-column formatting here, use the command \onecolumngrid
% Number subsections in the appendices as in the main text,
% except skip the capital Roman numerals.

\renewcommand{\thesection}{\Alph{section}}
\renewcommand{\thesubsection}{\Alph{section} \arabic{subsection}}
\renewcommand{\thesubsubsection}{\Alph{section} \arabic{subsection} \roman{subsubsection}}

% Label the equations in Appendix L as L1, L2, ...
\makeatletter\@addtoreset{equation}{section}
\def\theequation{\thesection\arabic{equation}}

\end{appendices}

%\bibliography{AQR}

%

%\printbibliography

\textbf{Data availability:} Supporting data are available in the figshare data repository (\url{https://doi.org/10.6084/m9.figshare.27089311.v1}). 

% Acknowledgements

\begin{acknowledgments}

This work received support from the Swedish Research Council (M.A.A. and S.G.); the Knut and Alice Wallenberg Foundation through the Wallenberg Center for Quantum Technology (WACQT) (C.C.-M. and S.G.); the European Union, Quantum Flagship project ASPECTS (grant agreement number 101080167) (M.A.A.) and ERC ESQuAT (grant number 101041744) (S.G.); the National Science Foundation, under QLCI grant OMA-2120757 (N.Y.H.) and grant number NSF PHY-1748958 (J.M.E.); the John Templeton Foundation (award number 62422) (J.A.M.G.); and NIST grant number 70NANB21H055 (J.A.M.G.). The studied device was fabricated in Myfab Chalmers, a nanofabrication laboratory.

\end{acknowledgments}

\section*{Author contributions}

N.Y.H. and S.G. conceived the experiment. M.A.A. J.A.M.G., J.M.E., N.Y.H. and S.G. performed the theoretical modelling and designed the experiment. M.A.A., P.J.S. and S.G. designed the device. C.C.-M. fabricated the device. M.A.A., P.J.S. and S.G. performed the experiments. M.A.A. and S.G. analyzed and interpreted the results. M.A.A., N.Y.H. and S.G. wrote the manuscript with feedback from P.J.S. and J.A.M.G. 

% \begin{thebibliography}{99}
% \end{thebibliography}

\clearpage
\onecolumngrid
\appendix

\renewcommand\thefigure{S\arabic{figure}}    
\renewcommand\thetable{S\arabic{table}}
\renewcommand\theequation{S\arabic{equation}}    
\setcounter{figure}{0} 
\setcounter{equation}{0}
\setcounter{table}{0}  

% Add your supplementary information document below

\newpage

%\tableofcontents

\section*{Supplementary Information}

\subsection{Full experimental setup and parameters}

\begin{figure*}[h]
\includegraphics[width=1\linewidth]{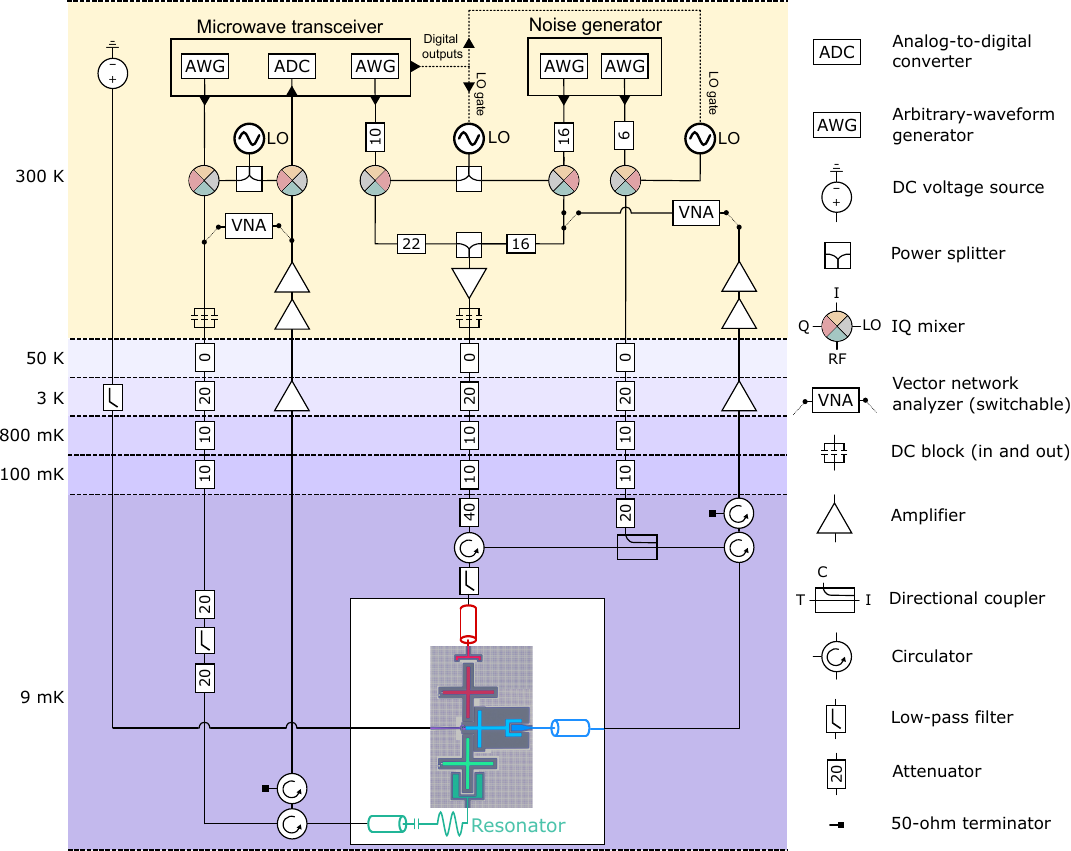}
\caption{Complete experimental set-up. See text for description. The in-phase-quadrature (IQ) mixers' ports,~I and Q, are both used to connect to the microwave transceiver. However, only port~I is shown to be wired; port~Q is omitted for clarity. Digital outputs from the microwave transceiver are used to gate the local oscillators (LO) to produce pulses of thermal microwave modes. The directional coupler has 3 ports, called the input port (I), coupled port (C), and through port (T).}\label{FigS1}
\end{figure*}

Figure~\ref{FigS1} shows a schematic of the experimental set-up used to study our quantum absorption refrigerator (QAR). The QAR is packaged in a copper enclosure and mounted on the mixing-chamber stage of a dilution refrigerator that reaches 10~mK. The QAR is packaged in multiple layers: two nested copper enclosures shield the QAR from electromagnetic waves. A $\mu$-metal enclosure protects the QAR from low-frequency magnetic fields. Microwave fields (both coherent and thermal) are routed to the QAR through highly attenuated input coaxial lines. The outgoing fields are routed through the output lines and are boosted by a cryogenic high-electron-mobility transistor (HEMT) amplifier (provided by Low Noise Factory) at 3~K and by room-temperature amplifiers. Microwave circulators separate the input and output signals. The resonator, dispersively coupled to qubit $Q_3$, is probed by a microwave feedline in reflection mode. We probe and drive $Q_1$ and $Q_2$, via the waveguide coupled to each, using two attenuated coaxial lines. We use a microwave directional coupler with three ports (input port, coupled port, and through port). Their arrangement allows us to use one output line for both $Q_1$'s and $Q_2$'s waveguides, taking advantage of the relevant transitions' nonoverlapping frequencies. Specifically, we combine the outgoing field from $Q_1$ (entering the through port) with the incoming field (entering the coupled port) intended for $Q_2$. The resultant field (exiting from the input port of the directional coupler) is routed by a circulator to $Q_2$'s waveguide.

We use a microwave transceiver (Vivace from Intermodulation Products, Sweden), with in-phase-quadrature (IQ) mixers and local oscillators (LO), for driving the qudits and for readout measurements of $Q_3$. The physical connections can be switched to a vector network analyzer (VNA). The VNA is used for continuous-wave spectroscopy that allows for basic characterization, e.g., of a qudit's frequency and anharmonicity. 

The waveguide coupled to $Q_1$ is populated with synthesized thermal microwave modes limited to a 50~MHz bandwidth centered at the qubit's frequency. An analogous statement concerns the waveguide coupled to $Q_2$. To populate the waveguides, we first continuously generate voltage noise with a white (flat) spectral density, using an arbitrary-waveform generator (AWG) limited to a 50~MHz bandwidth centered at a frequency of 100~MHz. The equipment used is a Keysight 3202A, which is limited to a 500~MHz bandwidth. Next, the generated noise is up-converted with IQ mixers and microwave tones ($\approx$ 4~GHz to 6~GHz) generated by an LO. Finally, the resulting continuous thermal radiation is chopped into pulses synchronized with $Q_3$-state readouts (see Fig.~3 in the main text). This process is performed via modulation (gating) of the LO outputs (Anapico APMS20G-4-ULN), with help from the transceiver's digital logic outputs, over time intervals of 10~ns and longer.
To characterize the qudits (e.g., to determine their frequencies), we can switch the input/output lines to VNA. Furthermore, to observe the three-body interactions with coherent drives (see Fig.~2 in the main text), we can switch qubit $Q_1$'s input line to the transceiver's AWG\footnote{Certain equipment, instruments, software, or materials are identified in this paper to specify the experimental procedure adequately.  Such identification is not intended to imply the recommendation or endorsement of any product or service by NIST; nor is it intended to imply that the materials or equipment identified are necessarily the best available for the purpose.}.
Table~\ref{TableS1} shows the QAR's experimentally inferred parameters.

\begin{table}
\centering
    \begin{tabular}{ l c c }
    \hline
    Parameter & Symbol & Value \\
    \hline
    $Q_1$ mode frequency & $\omega_1/(2\pi)$ & 5.327 GHz\\ 
    $Q_2$ mode-tunable frequency range & $\omega_2/(2\pi)$ & 4.2~GHz to 4.9 GHz\\
    $Q_2$ resonant frequency & $\omega_2^{\rm res}/(2\pi)$ & 4.629 GHz\\
    $Q_3$ mode frequency & $\omega_3/(2\pi)$ & 3.725 GHz\\
    $Q_1$ anharmonicity & $\alpha_1/(2\pi)$ & -213.4~MHz\\
    $Q_2$ anharmonicity & $\alpha_2/(2\pi)$ & -205.1~MHz\\
    $Q_3$ anharmonicity & $\alpha_3/(2\pi)$ & -237.8~MHz\\
    $Q_1$ radiative-emission rate & $\Gamma_1/(2\pi)$ & 70 kHz\\
    $Q_2$ radiative-emission rate & $\Gamma_2/(2\pi)$ & 7.2 MHz\\
    $Q_3$ natural energy-relaxation time & $T_{\rm relax}$ & 16.8~$\mu$s\\
    Three-body-coupling rate & $A/(2\pi)$ & 3.2~MHz\\
    %$Q_3$ energy relaxation time &$T_1$ & 16.8~$\mu$s\\
    
    \hline
    \end{tabular}
    \caption{Experimentally measured parameters' values.}\label{TableS1}
    \end{table}

\subsection{Theoretical model}

The Hamiltonian of our QAR's three-qudit system can be written as
\begin{equation}
\hat H = \sum_{i=1}^3 
\left(\tilde\omega_i \hat {\tilde a}^\dagger_i \hat {\tilde a}_i + \alpha_i \hat {\tilde a}^\dagger_i \hat {\tilde a}^\dagger_i \hat {\tilde a}_i \hat {\tilde a}_i/2 \right) 
+ g_{12} \left(\hat {\tilde a}^\dagger_1 \hat {\tilde a}_2 + \hat {\tilde a}_1 \hat {\tilde a}^\dagger_2 \right) 
+ g_{23} \left(\hat {\tilde a}^\dagger_2 \hat {\tilde a}_3 + \hat {\tilde a}_2 \hat {\tilde a}^\dagger_3 \right) \, .
\label{eq:MainHam}
\end{equation}
$\hat {\tilde a}_i$ and $\hat {\tilde a}^\dagger_i$ denote qudit $Q_i$'s annihilation and creation operators. $\tilde\omega_i$ and $\alpha_i$ denote qudit $Q_i$'s bare mode frequency and anharmonicity. $g_{12}$ ($g_{23}$) denotes the rate of the coupling between qudits $Q_1$ and $Q_2$ (qudits $Q_2$ and $Q_3$). 

We engineer an effective three-body interaction by meeting the resonance condition $\omega_1 + \omega_3 = 2\omega_2 + \alpha_2$ and using Josephson junctions that facilitate the four-wave mixing~\cite{ren2020}. The three-body interaction interchanges the three-qudit states $\ket{101}$ and $\ket{020}$ (see the notation and Fig.~1c in the main text). 

We now introduce an approximation to $\hat H$.
Denote by $A$ the rate at which $\ket{101}$ and $\ket{020}$ couple coherently.
$\hat H$ is well-approximated by the effective Hamiltonian
\begin{equation}
\hat H_{\rm eff} = \sum_{i=1}^3 
 \omega_i \hat a^\dagger_i \hat a_i + \sum_{i, j = 1}^3 \frac{\alpha_{ij}}{2} \hat a^\dagger_i \hat a_i \hat a^\dagger_j \hat a_j +
 A\left(\ket{101}\langle 020| + \ket{020}\langle 101|\right) .
\end{equation}\label{eq:SimHam}

\noindent The simplification follows from black-box quantization~\cite{nigg2012}
and second-order time-independent perturbation theory~\cite{ren2020}. $H_{\rm eff}$ depends on dressed modes associated with annihilation and creation operators $\hat a_i$ and $\hat a^\dagger_i$, dressed-mode frequencies $\omega_i$, and self-Kerr and cross-Kerr coupling rates $\alpha_{ij}$.

Let us introduce into the model a drive, as shown in Fig.~2 of the main text. This coherent drive is applied to $Q_1$ at the drive frequency $\omega_{\rm d}$ and the rate $\Omega$. Upon adding a drive term to $\hat H_{\rm eff}$, we apply the rotating-wave approximation. We obtain the full Hamiltonian
\begin{equation}
\hat H_{\rm D1} = \sum_{i=1}^3 
\delta_i \hat a^\dagger_i \hat a_i +
\sum_{i, j = 1}^3  \frac{\alpha_{ij}}{2} \hat a^\dagger_i \hat a_i \hat a^\dagger_j  \hat a_j  
+ A\left(\ket{101}\langle 020| + \ket{020}\langle 101|\right) + \frac{\Omega}{2} \left( \hat {\tilde a}_1 + \hat {\tilde a}^\dagger_1 \right).
\end{equation}\label{eq:DriveHam}
The effective frequency $\delta_i \coloneqq \omega_d - \omega_i$. 

To model the measurements of Fig.~2, we solve a Lindblad quantum master equation:
\begin{align}
    \frac{\partial \hat \rho}{\partial t} 
    & = -\frac{i}{\hbar}[\hat H_{D1},\rho] 
    + \sum_{i=1}^{3} \Gamma_i \mathcal{D}[\hat {\tilde a}_i, \rho] .
\end{align}\label{eq:MasterEqn}
\noindent We model the waveguides as zero-temperature baths (with average occupation numbers of 0). The qudits can interact with such baths just through spontaneous emission~\cite{carmichael1973}. We have suppressed the time dependence of $\rho$ in our notation. At all times $t$, $\rho$ approximately equals the tensor product of the qudits' reduced states, $\rho_i$: 
$\rho = \rho_{\rm 1} \otimes \rho_{\rm 2} \otimes \rho_{\rm 3}$. $\Gamma_{i = 1, 2}$ denotes the rate at which qudit $Q_i$ couples to its waveguide. $\Gamma_{E}$, $Q_3$'s natural energy-relaxation rate, is related to the qubit's energy-relaxation time $T_{\rm relax}$ when refrigerator is inactive through $\Gamma_{E} = 1/T_{\rm relax}$.
The dissipator superoperator $\mathcal{D}$ is defined through
\begin{equation}
\mathcal{D}[\hat A, \hat B] = \hat A \hat B \hat A^{\dagger} - \frac{1}{2}\left(\hat A^{\dagger} \hat A  \hat B + \hat B \hat A^{\dagger} \hat A\right) ,
\end{equation}\label{eq:DefDiss}
for operators $\hat A$ and $\hat B$.

To model the QAR interacting with two heat baths, we limit our analysis to a subspace of the full Hilbert space. This subspace is spanned by the basis $\mathcal{B} = \{\ket{000}, \ket{100}, \ket{010}, \ket{001}, \ket{002}, \ket{102}, \ket{020}, \ket{101} \}$ containing 8 levels.

The population of $\ket{i} \in \mathcal{B}$ is $p_i \coloneqq \rho_{ii} \coloneqq \bra{i} \hat \rho \ket{i}$. We formulate an 8-level population-rate model to calculate these populations's dynamics. To capture the coherent exchanges between $\ket{101}$ and $\ket{020}$, we also include an off-diagonal element $\rho_{\rm coh} = \bra{101} \hat \rho\ket{020}$ in the model. 
Combining these ingredients, we introduce the rate equation for the populations:
\begin{equation}
\frac{d}{dt} \begin{pmatrix}
p_{\ket{000}}(t) \\
p_{\ket{100}}(t) \\
p_{\ket{010}}(t) \\
p_{\ket{001}}(t) \\
p_{\ket{002}}(t) \\
p_{\ket{102}}(t) \\
p_{\ket{020}}(t) \\
p_{\ket{101}}(t) \\
\rho_{\rm coh}(t) \\

\end{pmatrix}
= R_{\rm P} \begin{pmatrix}
p_{\ket{000}}(t) \\
p_{\ket{100}}(t) \\
p_{\ket{010}}(t) \\
p_{\ket{001}}(t) \\
p_{\ket{002}}(t) \\
p_{\ket{102}}(t) \\
p_{\ket{020}}(t) \\
p_{\ket{101}}(t) \\
\rho_{\rm coh}(t)
\end{pmatrix} .
\label{eq:PopEqn}\end{equation}

The rate matrix $R_{\rm P}$ is defined as \\

\setlength{\arraycolsep}{1pt}
\begin{equation}
R_{\rm P} \coloneqq 
\footnotesize
\left(\begin{array}{ccccccccc}
-\Gamma_{1\uparrow}+\Gamma_{2\uparrow}+\Gamma_{3\uparrow} & \Gamma_{1\downarrow} & \Gamma_{2\downarrow} & \Gamma_{3\downarrow} & 0 & 0 & 0 & 0 & 0 \\
\Gamma_{1\uparrow} & -(\Gamma_{1\downarrow}+\Gamma_{3\uparrow}) & 0 & 0 & 0 & 0 & 0 & \Gamma_{3\downarrow} & 0 \\
\Gamma_{2\uparrow} & 0 & -(\Gamma_{2\downarrow}+2 \Gamma_{2\uparrow}) & 0 & 0 & 0 & 2 \Gamma_{2\downarrow} & 0 & 0 \\
\Gamma_{3\uparrow} & 0 & 0 & -(\Gamma_{3\downarrow}+\Gamma_{1\uparrow} + 2\Gamma_{3f\uparrow}) & 2\Gamma_{3f\downarrow} & 0 & 0 & \Gamma_{1\downarrow} & 0 \\
0 & 0 & 0 & 2\Gamma_{3f\uparrow}  & -(\Gamma_{1\uparrow} + 2\Gamma_{3f\downarrow}) & \Gamma_{1\downarrow} & 0 & 0 & 0 \\   
0 & 0 & 0 & 0 & \Gamma_{1\uparrow} & -( \Gamma_{1\downarrow} + 2\Gamma_{3f\downarrow}) & 0 & 2\Gamma_{3f\uparrow} & 0 \\
0 & 0 & 2 \Gamma_{2\uparrow} & 0 & 0 & 0 & - 2 \Gamma_{2\downarrow} & 0 & 2A \\
0 & \Gamma_{3\uparrow} & 0 & \Gamma_{1\uparrow} & 0 &  2\Gamma_{3f\downarrow} & 0 & -(\Gamma_{1\downarrow}+\Gamma_{3\downarrow} +  2\Gamma_{3f\uparrow}) & -2A \\
0 & 0 & 0 & 0 & 0 & 0 & -A & A & \Gamma_{\rm C}
\end{array}\right) %. 
\label{eq:PopMatrix}
\end{equation}

We define the $\Gamma$'s as follows. Denote by $n_i$ the average number of synthesized quasithermal photons in the waveguide coupled to qudit $Q_i$; and, by $n_{i, {\rm res}}$, the average number of native thermal (i.e., residual) photons already present in waveguide $i = 1, 2$. $n_{\rm{3, res}}$ is the effective average number of photons that populate qudit $Q_3$'s environment and that affect $Q_3$'s \emph{first transition}, $\ket{000} \leftrightarrow \ket{001}$. We also introduce $n_{3f, {\rm res}}$ as the average number of photons affecting the $Q_3$'s \emph{second transition}, $\ket{001} \leftrightarrow \ket{002}$.
In terms of these quantities, we define lowering rates $\Gamma_{i\downarrow} = \Gamma_i (n_i + n_{i, {\rm res}} + 1)$ and raising rates $\Gamma_{i\uparrow} = \Gamma_i (n_i + n_{i, {\rm res}})$ for qudits $i \in \{1, 2\}$, as well as $\Gamma_{3\downarrow} = \Gamma_{\rm E} (n_{3, {\rm res}} + 1)$ and $\Gamma_{3\uparrow} = \Gamma_{\rm E} n_{3, {\rm res}} $. Similarly, $\Gamma_{3f\uparrow} = \Gamma_{\rm E} n_{3f, {\rm res}}$ and $\Gamma_{3f\downarrow} = \Gamma_{\rm E} (n_{3f, {\rm res}} + 1)$ denote the analogous rates at the frequency of $Q_3$'s second transition ($\ket{001} \leftrightarrow \ket{002}$). For brevity, Eq.~\eqref{eq:PopMatrix} also contains a new notation $\Gamma_{\rm C}$ defined as

$\Gamma_{\rm C} = -(2 \Gamma_{2\downarrow} + 2 \Gamma_{2\uparrow} + \Gamma_{1\downarrow} + \Gamma_{1\uparrow} + \Gamma_{3\downarrow} + \Gamma_{3\uparrow})/4$. 

To model the data in Figs.~3b, ~4a and~4b, we solve Eq.~\eqref{eq:PopEqn} with the initial condition
\begin{multline}
\label{eq_vector}
\bm{(} p_{\ket{000}}(0), p_{\ket{100}}(0), p_{\ket{010}}(0), p_{\ket{001}}(0), p_{\ket{002}}(0), p_{\ket{102}}(0), p_{\ket{020}}(0), p_{\ket{101}}(0), \rho_{\rm coh}(0) \bm{)} \\ =
(0.028, 0.003, 0.007, 1 - 0.028 - 0.0008, 0.0008, 0, 0, 0, 0) .
\end{multline}

\noindent 

This vector represents the state prepared before the refrigeration---the state that the system is in after the $\pi$-pulse that impacts $Q_3$'s 0--1 transition (see Fig.~3b in the main text). Before the $\pi$-pulse, the $Q_3$ states $\ket{1}$ and $\ket{2}$ contain residual populations of $0.028$ (experimentally determined) and $0.028^2 = 0.0008$, respectively. We estimate $Q_3$'s $\ket{2}$ population under two assumptions: First, $Q_3$'s environment acts as a bath with the effective temperature $T_{\rm E}$. This temperature is the same for both the transitions, $\ket{0} \leftrightarrow \ket{1}$ and $\ket{1} \leftrightarrow \ket{2}$.
% 0--1 and 1--2. 
Second, Boltzmann factors determine the detailed-balance rate for both these transitions.
Therefore, $P_1  = P_0 \exp\left[-\hbar \omega_3/(k_\mathrm{B} T_{\rm E}) \right]$, and $P_2 = P_1 \exp\left[-\hbar (\omega_3 + \alpha_3)/(k_\mathrm{B} T_{\rm E}) \right] = P_0 \exp\left[-\hbar (2\omega_3 + \alpha_3)/(k_\mathrm{B} T_{\rm E}) \right] \approx \left\{ \exp\left[-\hbar \omega_3/(k_\mathrm{B} T_{\rm E}) \right] \right\}^2$. The last approximation is justified because $\alpha_3 \ll 2\omega_3$. Our assumption would have been undermined if two-level fluctuators were in $Q_3$'s environment, which could lead to widely different $T_{\rm E}$ across short frequency ranges~\cite{kulikov2020}. However, we have observed that our measured $P_{\rm res}$ does not lie outside a fixed narrow range (see Sec.~VII below) across thermal cyclings of the device. This observation supports our assumption that $T_{\rm E}$ is uniform across small frequency ranges.
Thus, $Q_3$'s $\ket{0}$ level has a population $1 - 0.028 - 0.0008$. The populations of $Q_1$'s state $\ket{1}$ is 0.003. The populations of $Q_2$'s level $\ket{1}$ is 0.007. The aforementioned $\pi$-pulse exchanges the populations of the $Q_3$ levels $\ket{0}$ and $\ket{1}$, yielding the vector~\eqref{eq_vector}.

Having introduced our dynamical model, we review the steady-state coefficient of performance (COP), following Ref.~\cite{hofer2016a}. Denote by $\mathcal{J}_T$ the steady-state current of heat drawn from the target system. Denote by $\mathcal{J}_H$ the steady-state current of heat drawn by $Q_1$ from the hot bath. The currents' ratio equals an absorption refrigerator's steady-state COP:

\begin{equation}
    \text{COP} \coloneqq \mathcal{J}_T / \mathcal{J}_H \, .
\end{equation}
% We calculate the COP at the refrigerator's steady state. 
The steady-state heat current drawn from bath $i \in \{1, 2, 3 \}$ is 
\begin{equation}
   \mathcal{J}_i 
   = \operatorname{Tr} \bm{(} \hat H \mathcal{L}_i \rho(\infty) \bm{)} .
\end{equation}
The Lindbladian therein acts as 
$\mathcal{L}_i \hat \rho 
= \Gamma_i \left\{ \left(n_i + n_{i, {\rm res}} + 1 \right) \mathcal{D} [ \hat {\tilde a}_i, \hat \rho] 
+ (n_i + n_{i, {\rm res}} )
\mathcal{D}[ \hat {\tilde a}_i^{\dagger}, \hat \rho] \right\}$.

%\FloatBarrier % Ensure all figures up to this point are placed
\subsection{Rabi population-measurement scheme}

\begin{figure*}[h] 
\includegraphics[width=1\linewidth]{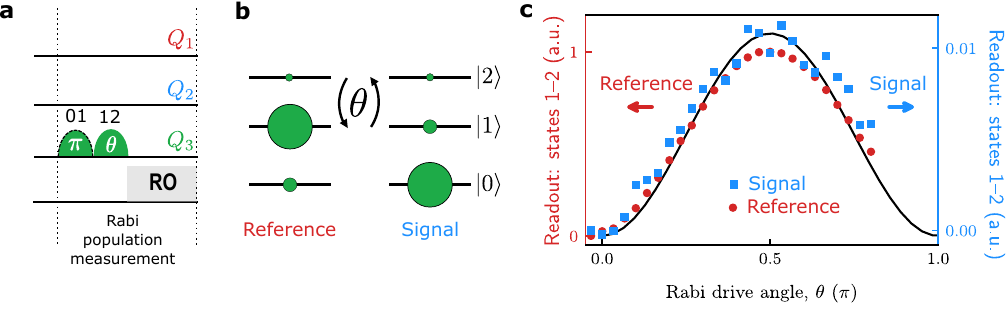}
\caption{Rabi measurement of the population in qubit $Q_3$'s $\ket{1}$ level. 
(a) Pulse sequence for the scheme. The notation 0--1 (1--2) indicates that the pulse drives the transition $|0\rangle \leftrightarrow |1\rangle$ ($|1\rangle \leftrightarrow |2\rangle$). $\theta$ denotes the pulse drive's Rabi angle. The pulse with the dashed outline is present in the reference measurement and absent from the signal measurement. The readout (RO) is performed in the subspace 
${\rm span} \{|1\rangle, |2\rangle \}$.  (b) Representation of the relative distribution of populations across qubit $Q_3$'s $|0\rangle$, $|1\rangle$, and $|2\rangle$ levels when $Q_3$ is in its natural steady-state. (c)~Rabi oscillations. The Rabi oscillation resulting from the 1--2 pulse after the 0--1 pulse provides a reference trace (red). The Rabi oscillation produced without the 1--2 pulse provides the signal---the actual population trace (blue).
}\label{FigS2}
\end{figure*}

Figure~\ref{FigS2} shows the pulse sequence used to measure $Q_3$'s level-$\ket{1}$ population more accurately than standard qubit readout allows~\cite{geerlings2013, jin2015a}. This Rabi scheme involves the $Q_3$ energy levels $|0\rangle$, $|1\rangle$, and $|2\rangle$. We measure the qubit-state readout in the subspace ${\rm span} \{|1\rangle, |2\rangle \}$. 

The measured readout voltage can be expressed as
\begin{equation}
   S = S_{0}P_{0} + S_{1}P_{1} + S_{2}P_{2} .
\end{equation}\label{eq:RO_voltage}
$P_{i}$ denotes the population of $Q_3$'s $i^{\rm th}$ state, for $i \in \{0, 1, 2\}$. $S_{i}$ represents the respective weighted coefficient. In the main text, we present the excited-state population, $P_1 + P_2$, as $P_{\rm exc}$.

The scheme for measuring $Q_3$'s excited-state population involves two different Rabi oscillations. We drive Rabi oscillations between $|1\rangle$ and $|2\rangle$ in each of two cases: with and without a prior $\pi$-pulse in the subspace ${\rm span} \{|0\rangle, |1\rangle \}$. The prior pulse interchanges the populations of $|0\rangle$ and $|1\rangle$, i.e., interchanges $P_{0}$ and $P_{1}$ (see the 0--1 pulse with the dashed outline in Fig.~\ref{FigS2}a). We measure and denote these Rabi oscillations' amplitudes by $S_\mathrm{R}$ (the Reference in Fig.~S2c) and $S_\mathrm{S}$ (the Signal in Fig.~S2c), respectively. We determine $S_\mathrm{R}$ and $S_\mathrm{S}$, while saving experiment runtime as follows: We perform a two-point measurement at $\theta = 0$ and $\theta = \pi$, to determine the amplitudes of the Rabi oscillations. 
From those measurements, we compute a ratio $S_{\rm P}$ related to the populations $P_{1}$ and $P_{2}$:
\begin{equation}
   S_\mathrm{P} = \frac{S_\mathrm{S}}{S_\mathrm{S} + S_\mathrm{R}} = \frac{P_{1} - P_{2}}{1 - 3P_{2}} \approx P_{1} - P_{2}
\end{equation}\label{eq:Pop_full_exp}
\noindent 
The last approximation holds if $P_{2} \ll 1$. 

Importantly, refrigeration with a large $n_\mathrm{H}$ results in a population $P_{1}$ that is very low ($\lesssim 0.001$) and is comparable to $P_{2}$. 
Our scheme cannot separate these two populations' contributions to the measured $S_{\rm P}$. Nonetheless, our quantitative 8-level model can yield a theoretical $P_{2}$, which is multiplied by 2 and added to our measured $S_\mathrm{P}$ value. Thus, we obtain the excited-state population: $P_{\rm exc} = P_{1} + P_{2}$. The result agrees with the calculation (Fig.~S3). %This is reported with the change of notation to $P$ in the main text and the rest of the SI.

\begin{figure}[h]
\includegraphics[width=0.6\linewidth]{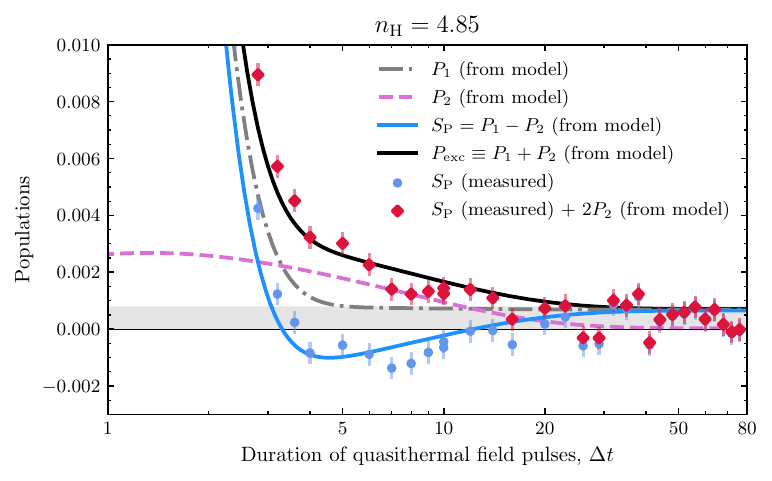}
\caption{Measured and calculated populations as functions of the time duration $\Delta t$ of the applied quasithermal fields, with $n_{\rm H} = 4.85$. This is one of the datasets presented in the main text's Fig.~3b. Recall that $P_1$ and $P_2$ denote the populations of $Q_1$'s $\ket{1}$ and $\ket{2}$ levels. These populations are calculated from the 8-level population-rate model in Eq.~\ref{eq:PopEqn}, presented in Sec.~II. The experimentally inferred ratio $S_{\rm P}$ is related to the levels' populations as $P_1 - P_2$. The measured data (blue circles) is compared to the blue curve calculated theoretically. From the measured $S_{\rm P}$ and calculated $P_2$, we obtain the excited-state population: $P_{\rm exc} = S_{\rm P} + 2P_2$ (cross-data points). This result is compared with the corresponding theoretical (black) curve, which represents $P_1 + P_2$. The error bars represent the standard deviations about the mean values, represented by symbols (see Sec.~IV for method). The shaded area represents estimate of the measurement's noise floor (see Sec.~IV for method). Accounting for $P_2$ is particularly important when $P_{\rm exc}$ reaches near 0.001 at intermediate $\Delta t$ values, when $P_{1}$ may become comparable to $P_{2}$. $P_{2}$ rises at intermediate $\Delta t$ because of the population transfer from the $Q_3$'s state $|1\rangle$ to state $|2\rangle$ at rate $2\Gamma_{3fu}$ (see the rate equation in sec.~II above). This process is impeded by the competing refrigeration process transferring population from state $|1\rangle$ to state $|0\rangle$ instead. At both low and high $\Delta t$ values, $P_2$ is negligible compared to $P_1$.
} \label{FigS3}
\end{figure}

\subsection{Population measurements near noise floor}

Our population measurement is particularly challenging when the population reaches very low values---specifically below 0.001, which appears to be the noise floor of our population measurements in Fig.~3b. Given this experimental limitation, we interpret the data near this noise floor with special consideration and draw conservative conclusions. Our measurements near the noise floor also yield negative values sometimes, as reported in~\cite{jin2015a, zhou2021}. The noise floor stems from the susceptibility of our measurement technique, which involves voltage-readout signals, to stochastic voltage noise in the measurement circuit---noise sourced predominantly by the circuit's amplifiers (see Fig.~S1).

We make a conservative estimate of the noise floor of our population measurements, shown in the main text's Fig.~3b, as follows. We consider three representative datasets for which $n_{\rm H} > 1$. We analyze fluctuations for values of time duration $\Delta t > 40~\mu$s. Here, our measurements appear close to or have reached the noise floor. Furthermore, the theoretical curves modelling the data relatively well, especially at low $\Delta t$, also indicate that the populations have reached their steady-state values for $\Delta t > 40~\mu$s . We compute the standard deviation for each curve separately, for $\Delta t > 40~\mu$s, yielding three estimates of the uncertainty in our measurements. We regard their average as the uncertainty in our measurements, $\sigma = 0.0004$. We identify the noise floor as $2\sigma = 0.0008$ above zero; it is shaded in gray in Fig.~3b. 

We perform a similar analysis to extract the value of, and uncertainty in, $P_{\rm SS}$ when it reaches its lowest value by refrigeration. Specifically, we analyze the experimental curve for $P_{\rm SS}$ as a function of $n_{\rm H}$, at the lowest $n_\mathrm{C} = 0.007$ (left plot of Fig.~4a). We acquired this dataset separately from the other three curves in Fig.~4a: We achieved a better signal-to-noise ratio (with more ensemble averaging and better qubit-readout settings), which is critical when $P_{\rm SS}$ is near the noise floor. We evaluate the lowest $P_{\rm SS}$ by computing the mean of the measured values of the curve that have declined below 0.0008 (an estimate of the noise floor identified above and depicted in Fig.~3b). This is a conservative evaluation of the mean. We are likely over-estimating the value of lowest $P_{\rm SS}$, thereby underestimating our refrigerator's capability. We estimate the uncertainty by computing the standard deviation of the same set of values of $P_{\rm SS}$. In conclusion, the mean and standard deviation are $3 \times 10^{-4}$ and $2 \times 10^{-4}$, respectively. A similar procedure was applied for other experimental curves in Fig.~4a to determine the uncertainty. For those curves, the standard deviation is $5 \times 10^{-4}$

\subsection{Engineering guidelines}
The main parameters usable to engineer our quantum refrigerator are the transition frequencies $\omega_1$, $\omega_2$, and $\omega_3$; the dissipation rates $\Gamma_1$ and $\Gamma_2$; and the three-body-interaction strength $A$. The interaction exchanges $\ket{101}$ and $\ket{020}$ only when the transition frequencies satisfy the resonance condition,
\begin{equation}
    \omega_1 +\omega_3 = 2\omega_2+\alpha_2.
\end{equation}
\noindent 

The greater the $A$, the more quickly our quantum refrigerator can cool the target qubit. In the perturbative approximation (when $g_{12}, g_{23}  \ll \omega_1, \omega_2, \omega_3$), $A$ has the form~\cite{ren2020}
\begin{equation}
    A = \sqrt{2} \, g_{12} \, g_{23}\left( \frac{1}{2\omega_2 + \alpha_2 - \omega_1} + \frac{1}{2\omega_2 + \alpha_2 - \omega_3}\right) .
\end{equation}
$A/(2\pi)$ is limited to values on the order of few MHz, for the nominal parameter values of $g_{12}, g_{23}, \omega_1, \omega_2, \omega_3$, within the applicability of the full Hamiltonian's rotating-wave approximation. %The steady-state performance of the refrigerator is characterized by the steady-state temperature and COP, which as discussed in the main text are primarily dependent on the 
Our quantum refrigerator has two steady-state metrics: the target's steady-state temperature, $T_{\rm SS}$, and the steady-state COP. We expect the generic quantum-absorption-refrigerator model~\cite{mitchison2019} to describe our quantum refrigerator. It implies 
\begin{equation}
T_{\rm SS} = \frac{2\omega_2+\alpha_2 - \omega_1}{(2\omega_2 +\alpha_2)/T_{\rm C} - \omega_1/T_{\rm H}} \, .
\end{equation}
\noindent This result relies on the assumption that the target qubit couples to its environment only weakly (at a low rate $\Gamma_{\rm E}$).

High $T_{\rm H}$ values (high $n_{\rm H}$ values) favor low $T_{\rm SS}$ values. A high $n_{\rm H}$ populates the $Q_1$'s level $\ket{1}$ more: $Q_1$ becomes excited at a rate $n_{\rm H}\Gamma_1$ and de-excites at a rate $(n_{\rm H} + 1)\Gamma_1$. Thus, $Q_1$'s $\ket{1}$ level achieves its greatest population
% the largest population $Q_1$'s state $\ket{1}$ is achieved 
when $n_{\rm H}$ is large. 
When $Q_1$'s $\ket{1}$ is highly populated, $\ket{101}$ is highly populated. This condition is the precursor to the cooling via the cascade process $\ket{101} \rightarrow \ket{020} \rightarrow \ket{010}$. 

For a different reason, low $T_{\rm C}$ values (low $n_{\rm C}$ values) favor low $T_{\rm SS}$ values. The levels $\ket{020}$ and $\ket{010}$ have low populations. The reverse process $\ket{010} \rightarrow \ket{020} \rightarrow \ket{101}$, which undermines the refrigeration, is therefore suppressed. 

For the sake of analyzing the refrigerator's dynamics, we model with an exponential function the decay of $Q_3$'s excited-state population, $P_{\rm exc}$, as a function of the duration $\Delta t$ for which the quasithermal fields are applied (see Fig~3b). The exponential approximation appears reasonable from our measurements, particularly until $P_{\rm exc}$ has declined to 0.01. We denote the cooling rate of $Q_3$ by $\Gamma_\mathrm{cool}$. $\Gamma_{\rm cool}$ is closely related to the inverse of reset time shown in Fig.~4b (the time needed for $Q_3$'s excited-state population to reach 0.01). In the optimization of $\Gamma_{\rm cool}$, the dissipation rates ($\Gamma_1$ and $\Gamma_2$) and $A$ play a critical role. Specifically, under the conditions $n_{\rm H} \gg 1$ and $n_{\rm C} \ll 1$, $n_{\rm H}\Gamma_1~(= \Gamma_{1\uparrow} \approx \Gamma_{1\downarrow})$ and $\Gamma_2 ~(\approx \Gamma_{2\downarrow})$ are the relevant rates. If these rates are too small, they become the limiting factors in the cascade process $\ket{001} \rightarrow \ket{101} \rightarrow \ket{020} \rightarrow \ket{010}$ that effects the refrigeration. On the other hand, if those rates are too high, the three-body coherent-exchange process suffers excessive dephasing and becomes the limiting factor. Figure~S4 shows a 2D plot of $\Gamma_{\rm cool}/A$ as a function of $n_{\rm H}\Gamma_1/A$ and $\Gamma_2/A$, calculated from our 8-level population model. Optimal values of $n_{\rm H}\Gamma_1/A$ and $\Gamma_2/A$ maximize $\Gamma_{\rm cool}$. Our experimental parameters fall at the gray cross-hair in Fig.~S4. The maximum $\Gamma_{\rm cool}$ corresponds to a minimum in reset (the time taken to reach $P_{\rm exc} = 0.01$). That is, Fig.~S4 guides the choice of parameters used to further reduce the reset time, as illustrated in Fig.~4b of the main text. A simpler version of the model, containing only 3 levels ($\ket{001}$, $\ket{101}$, and $\ket{020}$), yields the optimal values of the parameters close to that of the 8-level model. Thus, this simpler model suffices for engineering the refrigerator. However, the 3-level model does not fit our main results quantitatively.

\begin{figure}[h]
\includegraphics[width=0.5\linewidth]{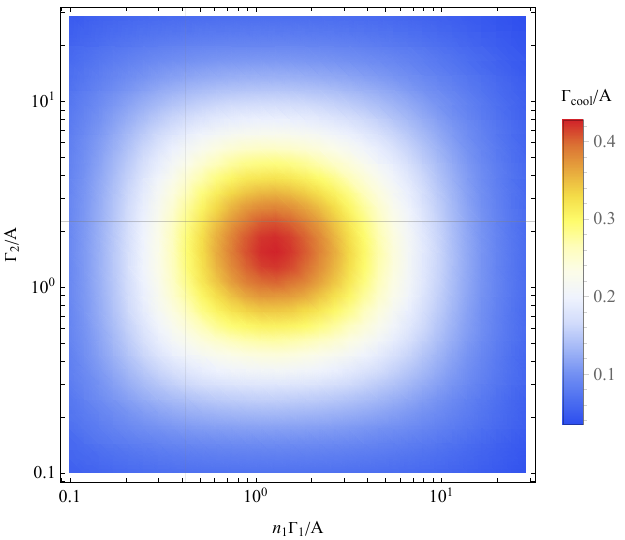}
\caption{Two-dimensional plot of $\Gamma_{\rm cool}/A$ as a function of $n_{\rm H}\Gamma_1/A$ and $\Gamma_2/A$, calculated from our 8-level model. $\Gamma_{\rm cool}$ is defined as the rate of the decay of $Q_3$'s excited-state population to $0.01$. The cross-hair indicates our experimentally realized values: $n_{\rm H} = 19.4$, $\Gamma_1/(2\pi) = 70$~kHz, $\Gamma_2/(2\pi) = 7.2$~MHz, and $A/(2\pi) = 3.2$~MHz. The experimentally determined $\Gamma_{\rm cool}/(2\pi) = 0.69$~MHz, which is close to the calculated $\Gamma_{\rm cool}/(2\pi) = 0.9$~MHz.
}
\label{FigS4}
\end{figure}

\subsection{Refrigeration's dependence on cold-bath temperature}

%\FloatBarrier % Ensure all figures up to this point are placed
\begin{figure}[h]
\includegraphics[width=0.8\linewidth]{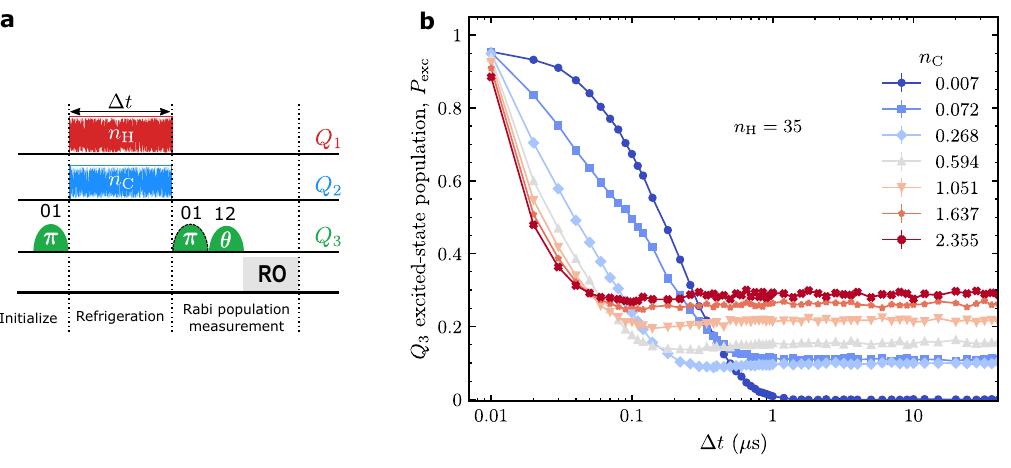}
\caption{Refrigeration's dependence on cold-bath temperature. (a) Pulse scheme. (b) $Q_3$'s excited-state population, $P_{\rm exc}$, as a function of the time duration $\Delta t$ of the application of quasithermal field, at select values of the average number $n_{\rm C}$ of thermal photons in the cold bath. The average number $n_{\rm H}$ of photons fixed at the high value~35.}\label{FigS5}
\end{figure}

Figure~\ref{FigS5} shows the $Q_3$'s excited-state population, $P_{\rm exc}$, as a function of the duration $\Delta t$ for which the quasithermal microwave modes are applied. We present these data at several values of the average number $n_{\rm C}$ of photons in the cold bath. The average number $n_{\rm H}$ of hot-bath thermal photons is a high value,~35.

Increasing $n_{\rm C}$ populates $Q_2$'s excited states incoherently, rendering the qutrit's state mixed~\cite{scigliuzzo2020}. The mixing hinders the three-body process $\ket{101} \rightarrow \ket{020}$. The refrigeration is impeded, so the steady-state $P_{\rm exc}$ increases.

\subsection{Refrigeration of $Q_3$ initiated in its natural steady state}

%\FloatBarrier % Ensure all figures up to this point are placed
\begin{figure}[h]
\includegraphics[width=0.8\linewidth]{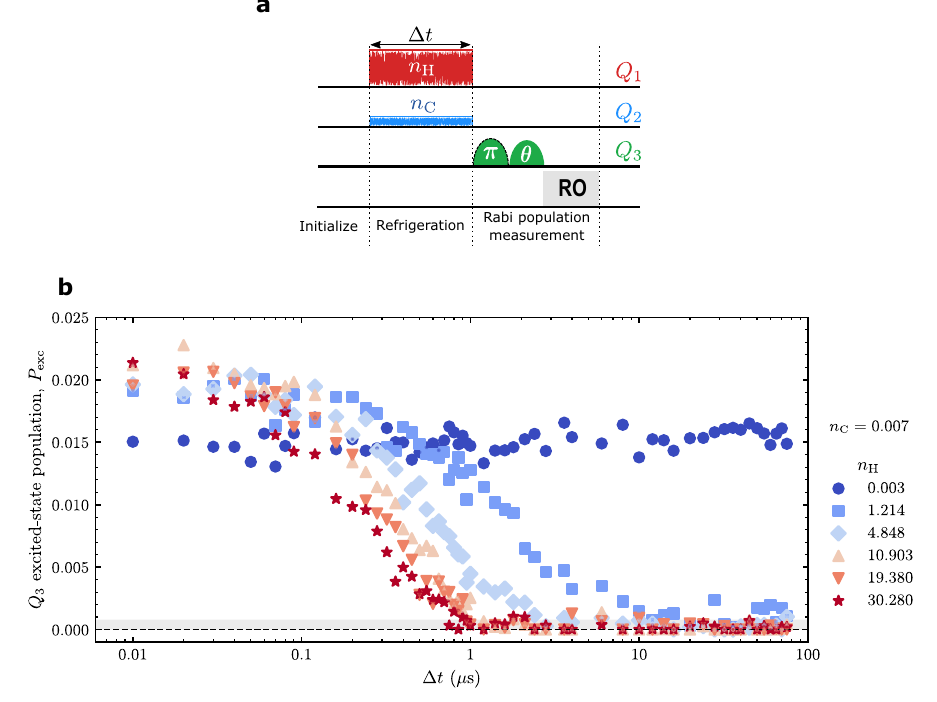}
\caption{Refrigeration of target qubit initialized with a residual $\ket{1}$ population of 0.020. (a) Pulse scheme. The qubit is not initialized entirely in the excited state. (b) $Q_3$'s excited-state population, $P_{\rm exc}$, as a function of the time duration $\Delta t$ of the thermal-microwave-mode pulse, for select values of the average number $n_{\rm H}$ of thermal photons in the hot bath.}\label{FigS6}
\end{figure}

In this study, we begin with $Q_3$ in the steady state that results from $Q_3$'s equilibrating as much as possible with its environment. (The state may, in fact, be far from equilibrium.) According to our measurements, $Q_3$'s $\ket{1}$ level has a population $P_{\rm exc}$ of $0.028$ in this state. We allow our quantum refrigerator to run for a duration $\Delta t$. Then, we perform a Rabi population measurement of $\ket{1}$. (See the pulse scheme in Fig.~\ref{FigS4}a.) $P_{\rm exc}$ is plotted as a function of $\Delta t$ in Fig.~\ref{FigS4}. Here, $P_{\rm exc}$ begins at a much lesser value than in the main text's Fig.~3b. Therefore, $P_{\rm exc}$ reaches its long-time value, here, by an earlier time $\approx 800$~ns.

We observed that $P_{\rm res}$ varied between 0.015 and 0.028 across long time frames (days) during the experiments. Such a large $P_{\rm res}$ (which is the motivation for reset) originates from uncontrolled non-equilibrium processes, such as stray infrared-radiation and inadequate device thermalization. Minute changes in the strength of these uncontrolled processes could lead to such changes in $P_{\rm res}$.

\end{document}